\documentclass{ws-p8-50x6-00}

\begin{document}

\title{Double parton scattering processes in $pp$ 
collisions and the scale factors.}

\author{A. Del Fabbro}

\address{Dipartimento di Fisica Teorica, Universit\`a di Trieste, Strada
Costiera 11, Miramare-Grignano and INFN, Sezione di Trieste,
I-34014 Trieste, Italy.\\ 
E-mail: delfabbr@ts.infn.it}


\maketitle

\abstracts{The double parton collisions become increasingly important at high
energies. The scale factors are the dimensional parameters which characterize
these interactions and are the physical observables that can provide a first
quantitative information on the correlations in the hadron structure. }

\section{Double Parton Collisions}
Multiple parton interaction processes, where different pairs  of partons
have an hard scattering in the same hadronic collision, become experimentally
feasible at high energies because of the growing flux of partons.
The hard partonic interactions occur independently  in different points inside
the overlap region of the two colliding hadrons and are separated by a
distance of the order of the hadron radius.
These processes are  characterized by two very different scales: the
hadronic dimension and the inverse of the hard scale in the partonic 
subprocesses. Therefore the interference effects betweeen the partonic
interactions  can be neglected and the expression of the cross sections
for these processes have a simple probabilistic interpretation.\\
In the case of double parton collisions, for the process $pp\rightarrow AB$, 
the cross section is 
\begin{eqnarray}
\sigma_D(AB)&=&\frac{m}{2}\int dx_1dx_1'dx_2dx_2'd^2b\nonumber\\
&\times& \Gamma(x_1,x_2;b)
\hat{\sigma}^A(x_1,x_1')
\hat{\sigma}^B(x_2,x_2')\Gamma'(x_1',x_2';b)\,
\label{sigmad2}
\end{eqnarray}
The non-perturbative input in this expression 
is the two-body parton distribution $\Gamma(x_1,x_2;b)$ where $x_1$ and $x_2$ 
are the momentum fractions and $b$ is the distance of the two partons in
transverce space.
  $\hat{\sigma}^A$ and $\hat{\sigma}^B$  are the partonic cross
 sections for the production of the final states $A$ and $B$ respectively
and $m=1 (m=2)$ when $A$ and $B$ are indistinguishable (distinguishable)
processes.

If one assumes that the the momentum fractions are not correlated and the
dependence on $b$ can be factorized,
the inclusive double parton
scattering cross
section for the two parton processes $A$ and $B$ reduces  to the
 simplest expression
\begin{equation}
\sigma_D=m\frac{\sigma_A \sigma_B}{2\sigma_{eff}}
\label{fact}
\end{equation}
where $\sigma_A$ and $\sigma_B$ are the inclusive single scattering 
cross sections for producing the final states $A$ and $B$ respectively.
The information on the
structure of the hadron in transverse space
 is summarized in the value of the scale
factor $\sigma_{eff}$ which is universal i.e. does not depend on the
considered process. The experimental value quoted by CDF is
$\sigma_{eff}=14.5\pm1.7^{+1.7}_{-2.3}$ mb \cite{cdf}.

Using this simple factorized expression one find that  
the double parton collisions give important effects
in several processes at the LHC, for example in the production of
$(W,Z)b{\bar b}$, $W+jets$, four $jets$, $b{\bar b}b{\bar b}$, etc ~\cite{yellow}.
An important  case is represented by the production
of $W b\bar b$ in a double parton scattering, which is a large source of
background in the Higgs boson production mechanism 
$pp\rightarrow W(H\rightarrow b\bar b)$
~\cite{higgs}. 

\section{Scale Factors}
The scale factor $\sigma_{eff}$ is related 
to the transverse size of the region where the 
hard interactions take place. More precisely, this parameter is  given
by the overlap in transverse space of the matter distribution of the two
interacting hadrons. 
 The measured  value of the scale factor is
sizably smaller than the naive expectations and it is an indication of
important correlations effects in the hadron structure~\cite{ctcorr}. 
In fact, while correlations in the momentum fractions will affect the two-body 
parton distributions   only for large values of $x$, the correlations in the
transverse parton coordinates are  always present due to the 
binding force. In a recent paper~\cite{ctmod} the
small value of the effective cross section  was explained by
 a simple correlated model of  the partonic 
population in the proton.  
In this model the distributions in transverse space 
of gluons and sea quarks are correlated with the configuration taken by the
valence quarks, in such a way that extended configurations of the valence give rise to a
large number of partons while compact configurations give rise to a smaller
number, as one expects for the growth of the energy of the gluonic field 
with the distance between the valence quarks.
On the other hand in the model the momentum fractions of the initial 
parton  pairs are uncorrelated.
One of the main features  is that there are two 
different kinds of partons as far as the distributions  in transverse space 
are concerned:  the valence quarks and the sea quarks and gluons.
As a consequence 
 the scale factor in a reaction involving valence quarks will be
different than the scale factor in a reaction involving sea quarks and gluons
~\cite{adfscale}.

The inclusive double parton scattering cross section is therefore expressed as 
\begin{eqnarray}
\sigma_D(A,B)&=&\frac{m}{2}\sum_{ijkl}\int dx_1dx_1'dx_2dx_2'd^2b \nonumber\\
&\times &\Gamma_{ij}(x_1,x_2;b)
\hat{\sigma}_{ik}^A(x_1,x_1')
\hat{\sigma}_{jl}^B(x_2,x_2')\Gamma_{kl}'(x_1',x_2';b)
\label{sigmad2}
\end{eqnarray}
where the indices $i$ and $j$ refer to the different kinds of
 partons, while 
dependence of $\Gamma$ on $x_1$, $x_2$ and $b$ is factorized:
$\Gamma_{ij}(x_1,x_2;b)=G_i(x_1)G_j(x_2)F_j^i(b)$,
and $G_i(x)$ are the usual one-body parton distributions.

The model introduces
the geometrical coefficients ~\cite{adfscale}
\begin{equation}
\Theta^{ij}_{kl}=\int d^2bF_k^i(b){F_l^j}'(b)\, ,
\label{fb}
\end{equation}
 wich have the dimension 
of the inverse of a cross section, so that   
the double scattering cross section is given by
\begin{equation}
\sigma_D(A,B)=\frac{m}{2}\sum_{ijkl}\Theta^{ij}_{kl}\sigma_{ij}(A)\sigma_{kl}(B)
\label{sigmad1}
\end{equation}
where $\sigma_{ij}(A)$ is the inclusive cross section for the process 
$ij\rightarrow A$.

By using the  equation         
 \begin{equation}
\sum_{ijkl}\Theta^{ij}_{kl}\sigma_{ij}(A)\sigma_{kl}(B)=
\frac{\sigma(A)\sigma(B)}{\sigma_{eff}}
\label{seff}
\end{equation}
we can therefore  obtain the scale factors, as a function of the c.m. energy,
 in various reactions 
where the rate of  double parton scatterings is large enough to measure
the corresponding scale factor. We have considered the following  production 
processes:   equal sign $WW$ ~\cite{stirl}, $Wjj$, $Wb{\bar b}$, 4-jets and ${b\bar b}{b\bar b}$. 
\begin{figure}
\centerline{
\epsfysize=8cm \epsfbox{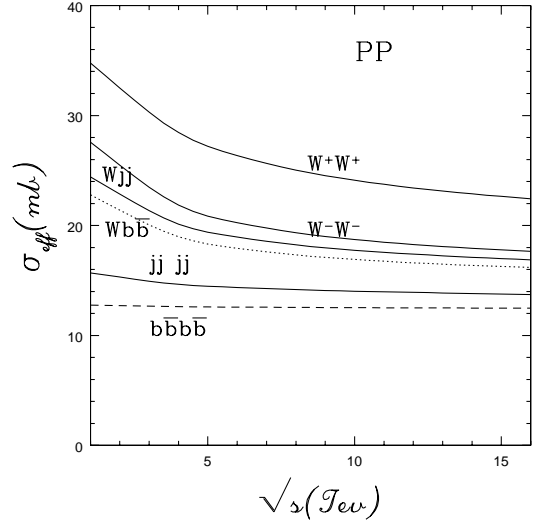}}
\caption[ ]{$\sigma_{eff}$ as a function of the c.m. energy in various
processes  in $pp$ collisions.}
\label{lhc}
\end{figure}
Our  results are plotted  in Fig.1,  
and show  strong difference between the scale factors
in the channels  with and without $W$ bosons,
as a consequence of the different role played by the  valence in the two
cases.  
The scale factor in  reactions involving valence quarks is in fact 
larger than that  in  reactions involving gluons and sea quarks.
This is due to the fact that 
in the model  the average transverse distance of a pair of 
valence quarks is larger than the average transverse distance 
 of a pair of gluons or sea quarks.    
The overall qualitative   
 indication of our calculations 
is that the scale factors are process and energy
dependent and that  they are the physical observables wich 
can provide  a
first quantitative information on the correlations in transverse space of the 
initial state partons.
\section*{Acknowledgments}
This work was partially supported by the Italian Ministry of University and of
Scientific and Technological Research by means of the Fondi per la Ricerca
scientifica-Universit\`a di Trieste.    



\end{document}